\documentclass[aps,prl,twocolumn,groupedaddress,floatfix]{revtex4-1}

\usepackage{graphicx}
\usepackage{natbib}

\begin{document}


\title{Evidence for a Universal Scaling of Length, Time and Energy in the Cuprate High Temperature Superconductors}


\author{J.D. Rameau}
\author{Z.-H. Pan}
\author{H.-B. Yang}
\author{G.D. Gu}
\author{P.D. Johnson}
\affiliation{Brookhaven National Laboratory, Upton, NY 11973, USA}

\date{\today}

\begin{abstract}
A microscopic scaling relation linking the normal and superconducting states of the cuprates in the presence of a pseudogap is presented using Angle Resolved Photoemission Spectroscopy. This scaling relation, complementary to the bulk universal scaling relation embodied by Homes' law, explicitly connects the momentum dependent amplitude of the d-wave superconducting order parameter at T$\sim 0$ to quasiparticle scattering mechanisms operative at T$\gtrsim T_{c}$. The form of the scaling is proposed to be a consequence of the Marginal Fermi Liquid phenomenology and the inherently strong dissipation of the normal pseudogap state of the cuprates.
\end{abstract}

\pacs{}

\maketitle

Not long after the discovery of high temperature superconductivity in the cuprates it was hypothesized that the transition temperature $T_{c}$ of these materials might be governed by the onset of phase coherence amongst ``preformed" Cooper pairs\cite{ref13, kivnat}. This scenario, essentially postulating a form of Bose condensation of such pairs, gives rise to a situation in which $T_{c}$ is lower than $T_{pair}$, the temperature at which the pairing amplitude of the superconducting order parameter develops. This point of view was bolstered early on by the observation of Uemura et al.\cite{uemura} that underdoped cuprates obey a seemingly universal scaling law, $\rho_{s0}\propto T_{c}$, where $\rho_{s0}$ is the superfluid density, or phase stiffness, at $T=0$, implying that the mechanism for high $T_{c}$ superconductivity does indeed entail a Bose condensation of well defined, preformed pairs rather than the traditional BCS mechanism in which the pairing amplitude of the order parameter and global phase coherence arise simultaneously. Recently however the Uemura relation was shown to be accompanied by another universal scaling law, ``Homes' law"\cite{ref1,ref2}, stating that $\rho_{s0}\propto \sigma_{DC}(T_{c})T_{c}$ where $\sigma_{DC}(T_{c})$ is the DC optical conductivity at $T\gtrsim T_{c}$. While Homes' law is valid over a much wider swath of the cuprate phase diagram than the Uemura relation, having been shown to apply to optimally and overdoped materials as well as the underdoped variety and even the new Fe base high $T_{c}$ superconductors\cite{fehomes}, a transparent picture of what it portends for the mechanism of high $T_{c}$ superconductivity in these materials has yet to emerge.

In this Report it is shown that angle resolved photoemission spectroscopy (ARPES) provides evidence for a complementary scaling relation between the momentum dependent single particle scattering rates of carriers at $T\gtrsim T_{c}$, at the Fermi energy $E_{F}$ on the Fermi surface (FS) 'arcs', and the magnitude of the superconducting gap at $T\sim0$ K, respectively. This finding, deriving from an examination of Homes' law\cite{ref1, ref2}, extends and clarifies the microscopic origins of that relationship, which was derived originally in the context of optical conductivity. As such, the present work represents a long sought after correlation between \emph{microscopic} spectral properties of the normal and superconducting states of high $T_{c}$ materials.

\begin{figure}
  \includegraphics[width= 76mm, bb = 0 58 489 738]{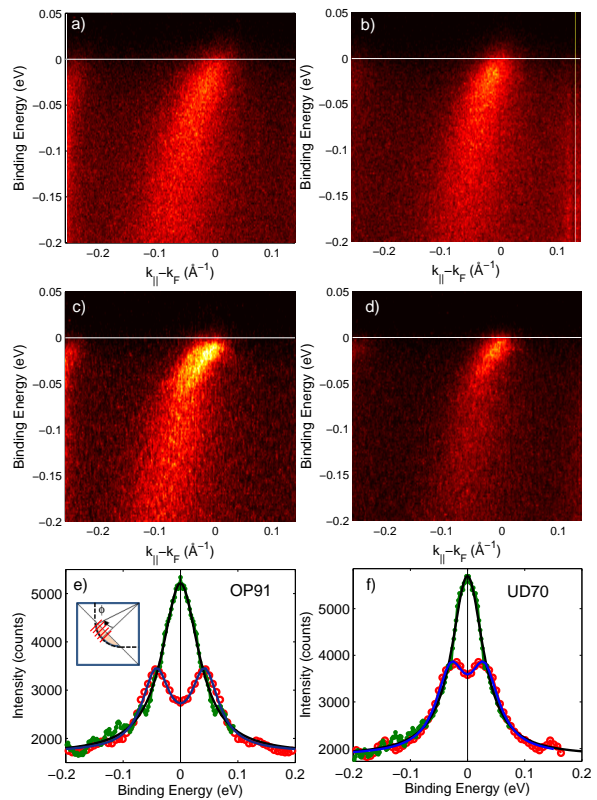}\\
  \caption{(Color online) Raw ARPES spectra of OP91 a) and UD70 b) at the tip of the Fermi arc, $\phi_{c}$, in the normal state just above $T_{c}$. Low temperature spectra ($T=15$ K, after deconvolution, are shown for the OP91 sample in c) and the UD70 sample in d). e) and f) show the $T_{c}$ and $T=15$ K symmetrized EDC's for OP91 and UD70, respectively, as well as Lorentzian fits thereof. The energy scale of the low temparture EDC's in panels e) and f) have been scaled up by a factor of $\pi$ to better illustrate Eq. \ref{eq1} evaluated at $\phi=\phi_{c}$ and the intensity scaled to half that of the normal state EDC's. The inset of panel e) illustrates the Fermi surface angle $\phi$ in relation to the normal state Fermi surface and the cuts represented by the current experiment (solid lines).}\label{NEWfig1}
\end{figure}

Homes' law has previously been interpreted as arising from a universal clean limit superconductivity ($\xi_{0}\ll\ell_{TC}$ where $\xi_{0}$ is Pippard's coherence length at $T=0$ and $\ell_{TC}$ is the electronic mean free path at $T\sim T_{c}$), universal dirty limit superconductivity \cite{ref2} ($\xi_{0}\geq\ell_{TC}$), ``hard core" boson scattering \cite{assa} and as indicative of normal state cuprates obeying a quantum critical-like relation of the form $k_{B}T_{c}\approx \hbar/\tau_{TC}$ where $\tau_{TC}$ is the mean free time (here in the sense of transport) of a normal state electron at $T\sim T_{c}$\cite{ref3}\cite{besthussey}. This ``Planckian" dissipation, viewable as a limit of the Marginal Fermi Liquid (MFL) phenomenology\cite{varma}, signifies that the observed electronic scattering is as rapid as is causally allowed. Separately, it has been suggested that Homes' law implies $\ell_{TC}\approx2\xi_{0}$\cite{tallon}. Altogether this implies the central issue in distinguishing various interpretations of Homes' law rests upon understanding the ratio $R=\ell_{TC}/\xi_{0}$, or equivalently, $R=\Delta_{0}\tau_{TC}$, a quantity often used to quantify the strength of scattering in a superconductor relative to the robustness of its pairing. To accomplish this in an ARPES experiment we must take full account of the d-wave nature of the superconducting order parameter and generalize from the coherence length and mean free path measured in transport to momentum dependent quantities $\xi_{0}(\phi)$ and $\ell_{TC}(\phi)$, $\phi$ being the FS angle as measured from the node (defined in the inset of Fig.\ref{NEWfig1}e). While such a generalization must be treated with care, especially near the node where $\xi_{0}(\phi)$ diverges, the result is nonetheless phenomenologically simple.

$\xi_{0}(\phi)$ can be measured in ARPES assuming a generalization of the coherence length, $\xi_{0}(\phi)=\frac{\hbar v_{F}^{0}(\phi)}{\pi\Delta_{0}(\phi)}$, where $v_{F}^{0}(\phi)$ and $\Delta_{0}(\phi)$ are the momentum dependent bare Fermi velocity at $T\sim T_{c}$ and the anisotropic superconducting gap at $T\sim0$, respectively. Similarly $\ell_{TC}(\phi)=1/\Delta k_{TC}(\phi)$ is the momentum dependent mean free path measured at $T\sim T_{c}$ with $\Delta k_{TC}(\phi)$ being the Lorentzian full width at half maximum of the momentum distribution curve (MDC) at $E=E_{F}$\cite{arpesmfl}. Noting that $\hbar v_{F}^{0}(\phi)\Delta k(\phi)=\hbar/\tau_{TC}(\phi)=\Gamma_{TC}(\phi)=2\Im\Sigma_{TC}$, where $\Im\Sigma$ is the imaginary part of the electron self energy, we find that the quantity of interest from the point of view of ARPES is $R(\phi)=\pi\Delta_{0}(\phi)/\Gamma_{TC}(\phi)$ where we recall that the inverse quasiparticle (QP) lifetime $\Gamma_{TC}(\phi)$ is the full width at half maximum of the peak in the ARPES energy distribution curve (EDC) line shape. Expressing $R(\phi)$ in terms of energy rather than length scales via the MDC equation has the advantage of obviating the need to infer the bare Fermi velocity from measurement.

It is evident from our definition of $R(\phi)$ that it can only have meaning when measured over the Fermi arc, understood to be the visible side of a nodal hole pocket\cite{ref16} as it exists at $T_{c}$, because $\Gamma_{TC}(\phi)$ is formally undefined for $\phi>\phi_{c}$ (where $\phi_{c}$ is the FS angle of the arc tip) due to the presence of the pseudogap (PG) at $E_{F}$ in optimal, underdoped and some lightly overdoped materials. Taken altogether the program for examining the quantity $R(\phi)$ using ARPES is to measure the QP lifetime at $E_{F}$ from the node to the tip of the Fermi arc, $\phi_{c}$, in the normal state at $T \gtrsim T_{c}$ (which are the states probed in transport experiments) and to compare these to measurements of the momentum dependent superconducting gap at the same points in k-space, for the same samples, at low temperature.

\begin{figure}
  \includegraphics[width = 86mm, bb = 13 249 595 544]{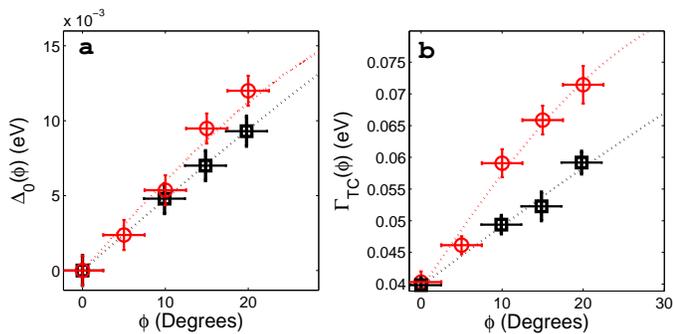}\\
  \caption{(Color online) a) Superconducting gap versus Fermi surface angle at T=15 K for UD70 (black squares) and OP91 (red circles). The gap is measured as the distance of the coherence peak in the superconducting state to $E_{F}$. b) Inverse lifetime at $T\sim T_{c}$ versus Fermi surface angle for UD70 (black squares) and OP91 (red circles). Fits described in the text are shown as dotted lines extrapolated towards the antinodal region\cite{ref8}. All error bars are derived from the fits.}\label{fig1}
\end{figure}

The experiment described above was carried out at beamline U13UB of the NSLS with a Scienta-2002 electron spectrometer. The end station was equipped with an open flow Helium cryostat. Two samples, high quality single crystals of Bi$_{2}$Sr$_{2}$CaCu$_{2}$O$_{8+x}$ (Bi2212) grown by the floating zone method, were used for measurements of $R(\phi)$: an optimally doped sample with $T_{c}$=91 K (OP91) and an underdoped sample with $T_{c}$=70 K (UD70). The lower $T_{c}$ of the UD70 sample was achieved by annealing in vacuum at 500 C for two days. Transition temperatures for both samples were ascertained prior to the ARPES measurement by SQUID magnetometry. The OP91 and UD70 samples were measured in their normal states at $T=95$ K and $T=70+$ K, respectively, and in their superconducting states at T=15 K. The overall resolution of the experiment was set to 12.5 meV in energy and $0.1^{o}$ in angle. The photon energies used were 18 eV (OP91) and 17.46 eV (UD70). All measurements were acquired within 48 hours of cleaving the samples at a chamber pressure of $1\times10^{-10}$ Torr.

\begin{figure}
  \includegraphics[width = 70mm,, bb = 38 176 553 597]{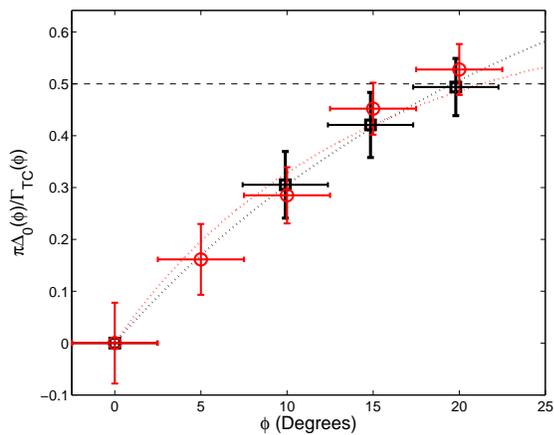}\\
  \caption{(Color online) a) Plot of $R(\phi)$ for OP91 (red circles) and UD70 Bi2212 (black squares). Error bars are formally propagated from those in Fig. \ref{fig1}. $R(\phi)$ obtained from the fits is superimposed on the data.}\label{fig2}
\end{figure}

Low temperature spectra used for acquiring $\Delta_{0}(\phi)$ were resolution corrected using the Lucy-Richardson method\cite{me}\cite{ref4}\cite{plumb} of deconvolution, yielding an effective energy resolution of ~4 meV. Values for the superconducting gap were determined by the binding energy of the coherence peak. Raw normal state and deconvolved low temperature data for $\phi=\phi_{c}$ are shown in Fig. \ref{NEWfig1}a)-d). Values of $\Delta_{0}(\phi)$ for UD70 and OP91 are presented in Fig. \ref{fig1}a along with pure d-wave ($\Delta(\phi)=\Delta_{0}^{AN}\sin(2\phi)$) fits, $\Delta_{0}^{AN}$ being determined by extrapolation from the nodal region\cite{ref8}. Normal state values for $\Gamma_{TC}(\phi)$ were acquired by fitting Lorentzians on a linear background to spectra symmetrized about $E_{F}$. Strictly speaking this procedure is only valid for states residing at $E_{F}$ and $k_{F}$ on the Fermi arc in the normal PG state of the copper oxides due to the presence of the previously observed particle-hole asymmetry in the nodal region\cite{ref4}. This is, by design, where the measurement is carried out. It has similarly been affirmed that the superconducting state is particle-hole symmetric well below $T_{c}$\cite{ref4}\cite{shen} so that symmetrization for our purposes is allowed at low temperature. The angular dependence of the inverse lifetimes, presented in Fig. \ref{fig1}b, were fit with the ``offset" d-wave \cite{besthussey} $\Gamma_{TC}(\phi)=\Gamma^{N}_{TC}+\delta\Gamma_{TC}\sin(2\phi)$ where $\delta\Gamma_{TC}=\Gamma_{TC}^{AN}-\Gamma_{TC}^{N}$, $\Gamma_{TC}^{N}$ and $\Gamma_{TC}^{AN}$ being the nodal and antinodal inverse lifetimes at $T_{c}$, respectively. $R(\phi)$, extracted from the data in Fig. \ref{fig1}, is plotted in Fig. \ref{fig2} along with analytical fits. We emphasize that $R(\phi)$ is obtained using total scattering rate (at each point in k-space) rather than just the anistropic component, as is appropriate to a comparison with the DC optical conductivity used in obtaining Homes' law.

\begin{figure}
  \includegraphics[width = 75mm,, bb = 65 178 525 594]{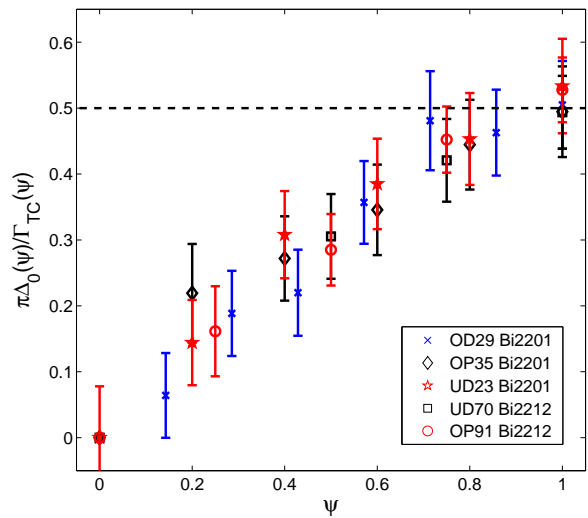}\\
  \caption{(Color online) Summary of available ARPES data from the current study as well as Ref. \cite{ref5} for $R(\psi)$. $\phi_{c}$ is taken in every case to simply be the last point to which a lifetime at the Fermi level could be reasonably ascertained. Error bars in the abscissa have been suppressed.}\label{fig3}
\end{figure}

To augment the present experimental results we reanalyzed data from a previous ARPES experiment performed on the single layer Bi2201 system under similar experimental conditions\cite{ref5}. In Fig. \ref{fig3}, which constitutes our main finding, the quantity $R(\psi)=\pi\Delta_{0}(\psi)/\Gamma_{TC}(\psi)$ versus the reduced Fermi surface angle $\psi\equiv\phi/\phi_{c}$ is plotted. Plotting $R$ in terms of $\psi$, which ranges between 0 at the node and 1 at the arc tip, rather than $\phi$, serves the purpose of collapsing data from samples with varying $T_{c}$ onto an equal footing. Remarkably, though the $T_{c}$'s of the materials thus investigated range between 25 K and 91 K, including single and bilayer systems, all materials are found to be very well approximated by a simple expression:
\begin{equation}\label{eq1}
\Delta_{0}(\psi)\cong\frac{\psi\Gamma_{TC}(\psi)}{2\pi}.
\end{equation}
While Eq. \ref{eq1} might be taken as a purely phenomenological expression it can be related to the important length scales of the system, allowing our previous derivation, such that
\begin{equation}\label{eq2}
\xi_{0}(\psi)\cong2\psi^{-1}\ell_{TC}(\psi).
\end{equation}
Eq. \ref{eq2} must be treated carefully in order to avoid divergence at the node.

Superconductors well described by the BCS theory have long been characterized as being in a ``clean" or ``dirty" limit based on comparisons of the type represented by Eqs. \ref{eq1} and \ref{eq2}\cite{degennes}. What such distinctions mean for superconductors possessing anisotropic order parameters is far from clear. In the present case those terms should evidently be eschewed because while the ratio of $\xi_{0}$ to $\ell_{TC}$, for example, is clearly a useful metric for parameterizing BCS superconductors there is no evidence of, or prescription for, a universal relationship between these quantities for a generic system as there is, say, between $T_{c}$ and $\Delta_{0}$. The present findings thus constitute evidence of a fundamental physical process in the cuprate superconductors that is not an obvious consequence of the BCS theory.

We postulate that a simple explanation for this behavior can be found by invoking the MFL phenomenology at $T_{c}$, $\Gamma_{TC}\propto T_{c}$, and the BCS superfluid, $\Delta_{0}\propto T_{c}$. If both of these properties hold across the Fermi arc then it is natural to conclude that $\Delta_{0}\propto \Gamma_{TC}$ will also hold across the fermi arc. Indeed, there is mounting evidence from ARPES and Scanning Tunneling Microscopy (STM) that $\Delta_{0}(\phi_{c})\propto T_{c}$ \cite{ref8}\cite{kurosawa}\cite{shen} and it was shown long ago that the MFL phenomenology is maintained in the ARPES spectrum of Bi2212\cite{arpesmfl}\cite{arpesmfl2}. Transport studies have also repeatedly reported observations of a correlation between the anisotropic dissipation of the normal state and $T_{c}$ \cite{besthussey}\cite{ref3}\cite{ref2} in the cuprates. It has also been shown recently that spin fluctuations, a leading candidate for the pairing mechanism of the curpates, leads (at least in some cases) to a T-linear scattering rate in the normal pseudogap state\cite{spinfluc}. Regardless, the physical content of Eq. \ref{eq1} is to imply that the interaction responsible for the anomalous non-Fermi liquid normal state scattering rate is intimately related to the interaction that gives rise to the pairing strength observed as a single particle gap on the Fermi arcs below $T_{c}$. It is hard to escape this conclusion given the ultimate proportionality between the superconducting gap at low $T$ and the imaginary part of the self energy at $T_{c}$ reported here.

The link between the pairing interaction and the electronic dissipation at $T_{c}$ has been remarked upon previously \cite{manifesto}, and indeed is explicitly predicted within the MFL phenomenology\cite{varma2}, though the existence of such a direct experimental relationship between the two, on the microscopic level, has not to our knowledge been previously reported. Eq. \ref{eq2} offers a more intuitive, real space picture of what the maximal dissipation of the normal state as a function of $T$ implies for superconductivity. Evidently, even if carriers were to experience a strong pairing interaction well above $T_{c}$, true pairs could not arise on the Fermi arc because the constituents rescatter before that information can be coherently propagated to a mate. $T_{c}$ appears to occur when all the carriers on the Fermi arc have a pairing amplitude \emph{and} can propagate that information, implying the relevance of an ``intra-pair" phase coherence to the magnitude of $T_{c}$. This length scale, $\xi_{0}$, set by the pairing strength plays a role fundamentally different than in BCS materials. If the scattering length of a single particle is too short relative to the size of the pairing potential it's in, it won't sense that potential. Such a dependence of the phase transition temperature $T_{c}$ purely on the relevant length and time scales of the system, rather than the details of the interaction, is the essence of quantum critical phenomena\cite{sachdev}. Eq. \ref{eq2} shows that in the cuprates, the new longer length scale is introduced a priori by the pairing interaction. Additionally, all carriers on the Fermi arc must be able to pair coherently before a gap can open, otherwise the symmetry of the d-wave order parameter would be violated.

Finally, we note that states at the Fermi arc tip appear to play a unique role in the phenomenology of the cuprates, on par with the high symmetry points of the nodal and antinodal states. There, Eqs. \ref{eq1} and \ref{eq2} reduce to $\Delta_{0}(\phi_{c})=\Gamma_{TC}(\phi_{c})/2\pi$ and $\xi_{0}(\phi_{c})=2\ell_{TC}(\phi_{c})$, respectively. This scaling, in relation to Eq. \ref{eq1}, is illustrated graphically in Fig. \ref{NEWfig1}e)-f). That PG states at higher momenta cannot satisfy this conditional relationship highlights a fundamental, if subtle, difference between the nodal and antinodal region of the Brillouin zone and suggests an inability of carriers tied up in the PG state above $T_{c}$ to ever condense into a true superfluid state.

Our findings evidence a universal, microscopic scaling relation between two fundamental properties of the cuprate superconductors: the normal state lifetime of carriers on the Fermi surface and the superconducting gap that arises from those states well below $T_{c}$. This relationship represents a clear departure from BCS theory by itself, yet suggests several key concepts of the BCS superfluid survive the quantum critical nature of the cuprates' anomalous normal state. The transition temperature is shown to be governed by a competition between length and time scales - pairing and single particle - both of which appear to be modulated by the same interaction. Lastly we note that the successful application of Homes' law to the pnictide and chalcogenide superconductors raises the intriguing possibility of performing ARPES experiments similar to those presented here in those systems.

\begin{acknowledgments}
We would like to acknowledge illuminating conversations with C.C. Homes, A.M. Tsvelik, M. Khodas and T.M. Rice. We would also like to acknowledge that the original inspiration for this work arose from conversations with Myron Strongin. This work is supported by the US DOE under Contract No. DE-AC02-98CH10886 and by the Center for Emergent Superconductivity, an Energy Frontier Research Consortium supported by the Office of Basic Energy Science of the Department of Energy.
\end{acknowledgments}



\begin{thebibliography}{26}%
\makeatletter
\providecommand \@ifxundefined [1]{%
 \@ifx{#1\undefined}
}%
\providecommand \@ifnum [1]{%
 \ifnum #1\expandafter \@firstoftwo
 \else \expandafter \@secondoftwo
 \fi
}%
\providecommand \@ifx [1]{%
 \ifx #1\expandafter \@firstoftwo
 \else \expandafter \@secondoftwo
 \fi
}%
\providecommand \natexlab [1]{#1}%
\providecommand \enquote  [1]{``#1''}%
\providecommand \bibnamefont  [1]{#1}%
\providecommand \bibfnamefont [1]{#1}%
\providecommand \citenamefont [1]{#1}%
\providecommand \href@noop [0]{\@secondoftwo}%
\providecommand \href [0]{\begingroup \@sanitize@url \@href}%
\providecommand \@href[1]{\@@startlink{#1}\@@href}%
\providecommand \@@href[1]{\endgroup#1\@@endlink}%
\providecommand \@sanitize@url [0]{\catcode `\\12\catcode `\$12\catcode
  `\&12\catcode `\#12\catcode `\^12\catcode `\_12\catcode `\%12\relax}%
\providecommand \@@startlink[1]{}%
\providecommand \@@endlink[0]{}%
\providecommand \url  [0]{\begingroup\@sanitize@url \@url }%
\providecommand \@url [1]{\endgroup\@href {#1}{\urlprefix }}%
\providecommand \urlprefix  [0]{URL }%
\providecommand \Eprint [0]{\href }%
\providecommand \doibase [0]{http://dx.doi.org/}%
\providecommand \selectlanguage [0]{\@gobble}%
\providecommand \bibinfo  [0]{\@secondoftwo}%
\providecommand \bibfield  [0]{\@secondoftwo}%
\providecommand \translation [1]{[#1]}%
\providecommand \BibitemOpen [0]{}%
\providecommand \bibitemStop [0]{}%
\providecommand \bibitemNoStop [0]{.\EOS\space}%
\providecommand \EOS [0]{\spacefactor3000\relax}%
\providecommand \BibitemShut  [1]{\csname bibitem#1\endcsname}%
\let\auto@bib@innerbib\@empty
\bibitem [{\citenamefont {Emery}\ and\ \citenamefont {Kivelson}(1995)}]{ref13}%
  \BibitemOpen
  \bibfield  {author} {\bibinfo {author} {\bibfnamefont {V.~J.}\ \bibnamefont
  {Emery}}\ and\ \bibinfo {author} {\bibfnamefont {S.~A.}\ \bibnamefont
  {Kivelson}},\ }\href@noop {} {\bibfield  {journal} {\bibinfo  {journal}
  {Phys. Rev. Lett.}\ }\textbf {\bibinfo {volume} {74}},\ \bibinfo {pages}
  {3253} (\bibinfo {year} {1995})}\BibitemShut {NoStop}%
\bibitem [{\citenamefont {Emergy}\ and\ \citenamefont
  {Kivelson}(1994)}]{kivnat}%
  \BibitemOpen
  \bibfield  {author} {\bibinfo {author} {\bibfnamefont {V.~J.}\ \bibnamefont
  {Emergy}}\ and\ \bibinfo {author} {\bibfnamefont {S.~A.}\ \bibnamefont
  {Kivelson}},\ }\href@noop {} {\bibfield  {journal} {\bibinfo  {journal}
  {Nature (London)}\ }\textbf {\bibinfo {volume} {374}},\ \bibinfo {pages}
  {434} (\bibinfo {year} {1994})}\BibitemShut {NoStop}%
\bibitem [{\citenamefont {Uemura}\ \emph {et~al.}(1989)\citenamefont {Uemura},
  \citenamefont {Luke}, \citenamefont {Sternlieb}, \citenamefont {Brewer},
  \citenamefont {Carolan}, \citenamefont {Hardy}, \citenamefont {Kadono},
  \citenamefont {Kempton}, \citenamefont {Kiefl}, \citenamefont {Kreitzman},
  \citenamefont {Mulhern}, \citenamefont {Riseman}, \citenamefont {Williams},
  \citenamefont {Yang}, \citenamefont {Uchida}, \citenamefont {Takagi},
  \citenamefont {Gopalakrishnan}, \citenamefont {Sleight}, \citenamefont
  {Subramanian}, \citenamefont {Chien}, \citenamefont {Cieplak}, \citenamefont
  {Xiao}, \citenamefont {Lee}, \citenamefont {Statt}, \citenamefont {Stronach},
  \citenamefont {Kossler},\ and\ \citenamefont {Yu}}]{uemura}%
  \BibitemOpen
  \bibfield  {author} {\bibinfo {author} {\bibfnamefont {Y.~J.}\ \bibnamefont
  {Uemura}}, \bibinfo {author} {\bibfnamefont {G.~M.}\ \bibnamefont {Luke}},
  \bibinfo {author} {\bibfnamefont {B.~J.}\ \bibnamefont {Sternlieb}}, \bibinfo
  {author} {\bibfnamefont {J.~H.}\ \bibnamefont {Brewer}}, \bibinfo {author}
  {\bibfnamefont {J.~F.}\ \bibnamefont {Carolan}}, \bibinfo {author}
  {\bibfnamefont {W.~N.}\ \bibnamefont {Hardy}}, \bibinfo {author}
  {\bibfnamefont {R.}~\bibnamefont {Kadono}}, \bibinfo {author} {\bibfnamefont
  {J.~R.}\ \bibnamefont {Kempton}}, \bibinfo {author} {\bibfnamefont {R.~F.}\
  \bibnamefont {Kiefl}}, \bibinfo {author} {\bibfnamefont {S.~R.}\ \bibnamefont
  {Kreitzman}}, \bibinfo {author} {\bibfnamefont {P.}~\bibnamefont {Mulhern}},
  \bibinfo {author} {\bibfnamefont {T.~M.}\ \bibnamefont {Riseman}}, \bibinfo
  {author} {\bibfnamefont {D.~L.}\ \bibnamefont {Williams}}, \bibinfo {author}
  {\bibfnamefont {B.~X.}\ \bibnamefont {Yang}}, \bibinfo {author}
  {\bibfnamefont {S.}~\bibnamefont {Uchida}}, \bibinfo {author} {\bibfnamefont
  {H.}~\bibnamefont {Takagi}}, \bibinfo {author} {\bibfnamefont
  {J.}~\bibnamefont {Gopalakrishnan}}, \bibinfo {author} {\bibfnamefont
  {A.~W.}\ \bibnamefont {Sleight}}, \bibinfo {author} {\bibfnamefont {M.~A.}\
  \bibnamefont {Subramanian}}, \bibinfo {author} {\bibfnamefont {C.~L.}\
  \bibnamefont {Chien}}, \bibinfo {author} {\bibfnamefont {M.~Z.}\ \bibnamefont
  {Cieplak}}, \bibinfo {author} {\bibfnamefont {G.}~\bibnamefont {Xiao}},
  \bibinfo {author} {\bibfnamefont {V.~Y.}\ \bibnamefont {Lee}}, \bibinfo
  {author} {\bibfnamefont {B.~W.}\ \bibnamefont {Statt}}, \bibinfo {author}
  {\bibfnamefont {C.~E.}\ \bibnamefont {Stronach}}, \bibinfo {author}
  {\bibfnamefont {W.~J.}\ \bibnamefont {Kossler}}, \ and\ \bibinfo {author}
  {\bibfnamefont {X.~H.}\ \bibnamefont {Yu}},\ }\href@noop {} {\bibfield
  {journal} {\bibinfo  {journal} {Phys. Rev. Lett.}\ }\textbf {\bibinfo
  {volume} {62}},\ \bibinfo {pages} {2317} (\bibinfo {year}
  {1989})}\BibitemShut {NoStop}%
\bibitem [{\citenamefont {Homes}\ \emph {et~al.}(2004)\citenamefont {Homes},
  \citenamefont {v.~Dordevic}, \citenamefont {Strongin}, \citenamefont {Bonn},
  \citenamefont {Liang}, \citenamefont {Hardy}, \citenamefont {Komiya},
  \citenamefont {Ando}, \citenamefont {Yu}, \citenamefont {Kaneko},
  \citenamefont {Zhao}, \citenamefont {Greven}, \citenamefont {Basov},\ and\
  \citenamefont {Timusk}}]{ref1}%
  \BibitemOpen
  \bibfield  {author} {\bibinfo {author} {\bibfnamefont {C.~C.}\ \bibnamefont
  {Homes}}, \bibinfo {author} {\bibfnamefont {S.}~\bibnamefont {v.~Dordevic}},
  \bibinfo {author} {\bibfnamefont {M.}~\bibnamefont {Strongin}}, \bibinfo
  {author} {\bibfnamefont {D.~A.}\ \bibnamefont {Bonn}}, \bibinfo {author}
  {\bibfnamefont {R.}~\bibnamefont {Liang}}, \bibinfo {author} {\bibfnamefont
  {W.~N.}\ \bibnamefont {Hardy}}, \bibinfo {author} {\bibfnamefont
  {S.}~\bibnamefont {Komiya}}, \bibinfo {author} {\bibfnamefont
  {Y.}~\bibnamefont {Ando}}, \bibinfo {author} {\bibfnamefont {G.}~\bibnamefont
  {Yu}}, \bibinfo {author} {\bibfnamefont {N.}~\bibnamefont {Kaneko}}, \bibinfo
  {author} {\bibfnamefont {X.}~\bibnamefont {Zhao}}, \bibinfo {author}
  {\bibfnamefont {M.}~\bibnamefont {Greven}}, \bibinfo {author} {\bibfnamefont
  {D.~N.}\ \bibnamefont {Basov}}, \ and\ \bibinfo {author} {\bibfnamefont
  {T.}~\bibnamefont {Timusk}},\ }\href@noop {} {\bibfield  {journal} {\bibinfo
  {journal} {Nature (London)}\ }\textbf {\bibinfo {volume} {430}},\ \bibinfo
  {pages} {539} (\bibinfo {year} {2004})}\BibitemShut {NoStop}%
\bibitem [{\citenamefont {Homes}\ \emph {et~al.}(2005)\citenamefont {Homes},
  \citenamefont {Dordevic}, \citenamefont {Valla},\ and\ \citenamefont
  {Strongin}}]{ref2}%
  \BibitemOpen
  \bibfield  {author} {\bibinfo {author} {\bibfnamefont {C.~C.}\ \bibnamefont
  {Homes}}, \bibinfo {author} {\bibfnamefont {S.~V.}\ \bibnamefont {Dordevic}},
  \bibinfo {author} {\bibfnamefont {T.}~\bibnamefont {Valla}}, \ and\ \bibinfo
  {author} {\bibfnamefont {M.}~\bibnamefont {Strongin}},\ }\href {\doibase
  10.1103/PhysRevB.72.134517} {\bibfield  {journal} {\bibinfo  {journal} {Phys.
  Rev. B}\ }\textbf {\bibinfo {volume} {72}},\ \bibinfo {pages} {134517}
  (\bibinfo {year} {2005})}\BibitemShut {NoStop}%
\bibitem [{\citenamefont {Homes}\ \emph {et~al.}(2010)\citenamefont {Homes},
  \citenamefont {Akrap}, \citenamefont {Wen}, \citenamefont {xu}, \citenamefont
  {nd~Q.~Li},\ and\ \citenamefont {Gu}}]{fehomes}%
  \BibitemOpen
  \bibfield  {author} {\bibinfo {author} {\bibfnamefont {C.~C.}\ \bibnamefont
  {Homes}}, \bibinfo {author} {\bibfnamefont {A.}~\bibnamefont {Akrap}},
  \bibinfo {author} {\bibfnamefont {J.~S.}\ \bibnamefont {Wen}}, \bibinfo
  {author} {\bibfnamefont {Z.~J.}\ \bibnamefont {xu}}, \bibinfo {author}
  {\bibfnamefont {Z.~W.~L.}\ \bibnamefont {nd~Q.~Li}}, \ and\ \bibinfo {author}
  {\bibfnamefont {G.~D.}\ \bibnamefont {Gu}},\ }\href@noop {} {\bibfield
  {journal} {\bibinfo  {journal} {Phys. Rev. B}\ }\textbf {\bibinfo {volume}
  {81}},\ \bibinfo {pages} {180508(R)} (\bibinfo {year} {2010})}\BibitemShut
  {NoStop}%
\bibitem [{\citenamefont {Lindner}\ and\ \citenamefont
  {Auerbach}(2010)}]{assa}%
  \BibitemOpen
  \bibfield  {author} {\bibinfo {author} {\bibfnamefont {N.~H.}\ \bibnamefont
  {Lindner}}\ and\ \bibinfo {author} {\bibfnamefont {A.}~\bibnamefont
  {Auerbach}},\ }\href@noop {} {\bibfield  {journal} {\bibinfo  {journal}
  {Phys. Rev. B}\ }\textbf {\bibinfo {volume} {81}},\ \bibinfo {pages} {054512}
  (\bibinfo {year} {2010})}\BibitemShut {NoStop}%
\bibitem [{\citenamefont {Zaanen}(2004)}]{ref3}%
  \BibitemOpen
  \bibfield  {author} {\bibinfo {author} {\bibfnamefont {J.}~\bibnamefont
  {Zaanen}},\ }\href@noop {} {\bibfield  {journal} {\bibinfo  {journal} {Nature
  (London)}\ }\textbf {\bibinfo {volume} {430}},\ \bibinfo {pages} {512}
  (\bibinfo {year} {2004})}\BibitemShut {NoStop}%
\bibitem [{\citenamefont {Abdel-Jawad}\ \emph {et~al.}(2006)\citenamefont
  {Abdel-Jawad}, \citenamefont {Kennett}, \citenamefont {Balicas},
  \citenamefont {Carrington}, \citenamefont {Mackenzie}, \citenamefont
  {McKenzie},\ and\ \citenamefont {Hussey}}]{besthussey}%
  \BibitemOpen
  \bibfield  {author} {\bibinfo {author} {\bibfnamefont {M.}~\bibnamefont
  {Abdel-Jawad}}, \bibinfo {author} {\bibfnamefont {M.~P.}\ \bibnamefont
  {Kennett}}, \bibinfo {author} {\bibfnamefont {L.}~\bibnamefont {Balicas}},
  \bibinfo {author} {\bibfnamefont {A.}~\bibnamefont {Carrington}}, \bibinfo
  {author} {\bibfnamefont {A.~P.}\ \bibnamefont {Mackenzie}}, \bibinfo {author}
  {\bibfnamefont {R.~H.}\ \bibnamefont {McKenzie}}, \ and\ \bibinfo {author}
  {\bibfnamefont {N.~E.}\ \bibnamefont {Hussey}},\ }\href@noop {} {\bibfield
  {journal} {\bibinfo  {journal} {Nature Physics}\ }\textbf {\bibinfo {volume}
  {2}},\ \bibinfo {pages} {821} (\bibinfo {year} {2006})}\BibitemShut {NoStop}%
\bibitem [{\citenamefont {Varma}\ \emph {et~al.}(1989)\citenamefont {Varma},
  \citenamefont {Littlewood},\ and\ \citenamefont {Schmitt-Rink}}]{varma}%
  \BibitemOpen
  \bibfield  {author} {\bibinfo {author} {\bibfnamefont {C.~M.}\ \bibnamefont
  {Varma}}, \bibinfo {author} {\bibfnamefont {P.~B.}\ \bibnamefont
  {Littlewood}}, \ and\ \bibinfo {author} {\bibfnamefont {S.}~\bibnamefont
  {Schmitt-Rink}},\ }\href@noop {} {\bibfield  {journal} {\bibinfo  {journal}
  {Phys. Rev. Lett.}\ }\textbf {\bibinfo {volume} {63}},\ \bibinfo {pages}
  {1996} (\bibinfo {year} {1989})}\BibitemShut {NoStop}%
\bibitem [{\citenamefont {Tallon}\ \emph {et~al.}(2006)\citenamefont {Tallon},
  \citenamefont {Cooper}, \citenamefont {Naqib},\ and\ \citenamefont
  {Loram}}]{tallon}%
  \BibitemOpen
  \bibfield  {author} {\bibinfo {author} {\bibfnamefont {J.~L.}\ \bibnamefont
  {Tallon}}, \bibinfo {author} {\bibfnamefont {J.~R.}\ \bibnamefont {Cooper}},
  \bibinfo {author} {\bibfnamefont {S.~H.}\ \bibnamefont {Naqib}}, \ and\
  \bibinfo {author} {\bibfnamefont {J.~W.}\ \bibnamefont {Loram}},\ }\href@noop
  {} {\bibfield  {journal} {\bibinfo  {journal} {Phys. Rev. B}\ }\textbf
  {\bibinfo {volume} {73}},\ \bibinfo {pages} {180504(R)} (\bibinfo {year}
  {2006})}\BibitemShut {NoStop}%
\bibitem [{\citenamefont {Valla}\ \emph {et~al.}(1999)\citenamefont {Valla},
  \citenamefont {Fedorov}, \citenamefont {Johnson}, \citenamefont {Wells},
  \citenamefont {Hulbert}, \citenamefont {Li}, \citenamefont {Gu},\ and\
  \citenamefont {Koshizuka}}]{arpesmfl}%
  \BibitemOpen
  \bibfield  {author} {\bibinfo {author} {\bibfnamefont {T.}~\bibnamefont
  {Valla}}, \bibinfo {author} {\bibfnamefont {A.~V.}\ \bibnamefont {Fedorov}},
  \bibinfo {author} {\bibfnamefont {P.~D.}\ \bibnamefont {Johnson}}, \bibinfo
  {author} {\bibfnamefont {B.~O.}\ \bibnamefont {Wells}}, \bibinfo {author}
  {\bibfnamefont {S.~L.}\ \bibnamefont {Hulbert}}, \bibinfo {author}
  {\bibfnamefont {Q.}~\bibnamefont {Li}}, \bibinfo {author} {\bibfnamefont
  {G.~D.}\ \bibnamefont {Gu}}, \ and\ \bibinfo {author} {\bibfnamefont
  {N.}~\bibnamefont {Koshizuka}},\ }\href@noop {} {\bibfield  {journal}
  {\bibinfo  {journal} {Science}\ }\textbf {\bibinfo {volume} {285}},\ \bibinfo
  {pages} {2110} (\bibinfo {year} {1999})}\BibitemShut {NoStop}%
\bibitem [{\citenamefont {Yang}\ \emph {et~al.}(2011)\citenamefont {Yang},
  \citenamefont {Rameau}, \citenamefont {Pan}, \citenamefont {Gu},
  \citenamefont {Johnson}, \citenamefont {Claus}, \citenamefont {Hinks},\ and\
  \citenamefont {Kidd}}]{ref16}%
  \BibitemOpen
  \bibfield  {author} {\bibinfo {author} {\bibfnamefont {H.-B.}\ \bibnamefont
  {Yang}}, \bibinfo {author} {\bibfnamefont {J.~D.}\ \bibnamefont {Rameau}},
  \bibinfo {author} {\bibfnamefont {Z.-H.}\ \bibnamefont {Pan}}, \bibinfo
  {author} {\bibfnamefont {G.~D.}\ \bibnamefont {Gu}}, \bibinfo {author}
  {\bibfnamefont {P.~D.}\ \bibnamefont {Johnson}}, \bibinfo {author}
  {\bibfnamefont {H.}~\bibnamefont {Claus}}, \bibinfo {author} {\bibfnamefont
  {D.~G.}\ \bibnamefont {Hinks}}, \ and\ \bibinfo {author} {\bibfnamefont
  {T.~E.}\ \bibnamefont {Kidd}},\ }\href@noop {} {\bibfield  {journal}
  {\bibinfo  {journal} {Phys. Rev. Lett.}\ }\textbf {\bibinfo {volume} {107}},\
  \bibinfo {pages} {047003} (\bibinfo {year} {2011})}\BibitemShut {NoStop}%
\bibitem [{\citenamefont {Pushp}\ \emph {et~al.}(2009)\citenamefont {Pushp}, ,
  \citenamefont {Parker}, \citenamefont {Pasupathy}, \citenamefont {Gomes},
  \citenamefont {Ono}, \citenamefont {Wen}, \citenamefont {Xu}, \citenamefont
  {Gu},\ and\ \citenamefont {Yazdani}}]{ref8}%
  \BibitemOpen
  \bibfield  {author} {\bibinfo {author} {\bibfnamefont {A.}~\bibnamefont
  {Pushp}}, , \bibinfo {author} {\bibfnamefont {C.~V.}\ \bibnamefont {Parker}},
  \bibinfo {author} {\bibfnamefont {A.~N.}\ \bibnamefont {Pasupathy}}, \bibinfo
  {author} {\bibfnamefont {K.~K.}\ \bibnamefont {Gomes}}, \bibinfo {author}
  {\bibfnamefont {S.}~\bibnamefont {Ono}}, \bibinfo {author} {\bibfnamefont
  {J.}~\bibnamefont {Wen}}, \bibinfo {author} {\bibfnamefont {Z.}~\bibnamefont
  {Xu}}, \bibinfo {author} {\bibfnamefont {G.}~\bibnamefont {Gu}}, \ and\
  \bibinfo {author} {\bibfnamefont {A.}~\bibnamefont {Yazdani}},\ }\href@noop
  {} {\bibfield  {journal} {\bibinfo  {journal} {Science}\ }\textbf {\bibinfo
  {volume} {324}},\ \bibinfo {pages} {1689} (\bibinfo {year}
  {2009})}\BibitemShut {NoStop}%
\bibitem [{\citenamefont {Rameau}\ \emph {et~al.}(2010)\citenamefont {Rameau},
  \citenamefont {Yang},\ and\ \citenamefont {Johnson}}]{me}%
  \BibitemOpen
  \bibfield  {author} {\bibinfo {author} {\bibfnamefont {J.~D.}\ \bibnamefont
  {Rameau}}, \bibinfo {author} {\bibfnamefont {H.-B.}\ \bibnamefont {Yang}}, \
  and\ \bibinfo {author} {\bibfnamefont {P.~D.}\ \bibnamefont {Johnson}},\
  }\href@noop {} {\bibfield  {journal} {\bibinfo  {journal} {J. Elec. Spec.
  Relat. Phenom.}\ }\textbf {\bibinfo {volume} {181}},\ \bibinfo {pages} {35}
  (\bibinfo {year} {2010})}\BibitemShut {NoStop}%
\bibitem [{\citenamefont {Yang}\ \emph {et~al.}(2008)\citenamefont {Yang},
  \citenamefont {Rameau}, \citenamefont {Johnson}, \citenamefont {Valla},
  \citenamefont {Tsvelik},\ and\ \citenamefont {Gu}}]{ref4}%
  \BibitemOpen
  \bibfield  {author} {\bibinfo {author} {\bibfnamefont {H.-B.}\ \bibnamefont
  {Yang}}, \bibinfo {author} {\bibfnamefont {J.~D.}\ \bibnamefont {Rameau}},
  \bibinfo {author} {\bibfnamefont {P.~D.}\ \bibnamefont {Johnson}}, \bibinfo
  {author} {\bibfnamefont {T.}~\bibnamefont {Valla}}, \bibinfo {author}
  {\bibfnamefont {A.}~\bibnamefont {Tsvelik}}, \ and\ \bibinfo {author}
  {\bibfnamefont {G.~D.}\ \bibnamefont {Gu}},\ }\href@noop {} {\bibfield
  {journal} {\bibinfo  {journal} {Nature (London)}\ }\textbf {\bibinfo {volume}
  {456}},\ \bibinfo {pages} {77} (\bibinfo {year} {2008})}\BibitemShut
  {NoStop}%
\bibitem [{\citenamefont {Plumb}\ \emph {et~al.}(2010)\citenamefont {Plumb},
  \citenamefont {Reber}, \citenamefont {Koralek}, \citenamefont {Sun},
  \citenamefont {Douglas}, \citenamefont {Aiura}, \citenamefont {Oka},
  \citenamefont {Eisaki},\ and\ \citenamefont {Dessau}}]{plumb}%
  \BibitemOpen
  \bibfield  {author} {\bibinfo {author} {\bibfnamefont {N.~C.}\ \bibnamefont
  {Plumb}}, \bibinfo {author} {\bibfnamefont {T.~J.}\ \bibnamefont {Reber}},
  \bibinfo {author} {\bibfnamefont {J.~D.}\ \bibnamefont {Koralek}}, \bibinfo
  {author} {\bibfnamefont {Z.}~\bibnamefont {Sun}}, \bibinfo {author}
  {\bibfnamefont {J.~F.}\ \bibnamefont {Douglas}}, \bibinfo {author}
  {\bibfnamefont {Y.}~\bibnamefont {Aiura}}, \bibinfo {author} {\bibfnamefont
  {K.}~\bibnamefont {Oka}}, \bibinfo {author} {\bibfnamefont {H.}~\bibnamefont
  {Eisaki}}, \ and\ \bibinfo {author} {\bibfnamefont {D.~S.}\ \bibnamefont
  {Dessau}},\ }\href@noop {} {\bibfield  {journal} {\bibinfo  {journal} {Phys.
  Rev. Lett.}\ }\textbf {\bibinfo {volume} {105}},\ \bibinfo {pages} {046402}
  (\bibinfo {year} {2010})}\BibitemShut {NoStop}%
\bibitem [{\citenamefont {Lee}\ \emph {et~al.}(2007)\citenamefont {Lee},
  \citenamefont {Vishik}, \citenamefont {Tanaka}, \citenamefont {Lu},
  \citenamefont {Sasagawa}, \citenamefont {Nagaosa}, \citenamefont {Devereaux},
  \citenamefont {Hussain},\ and\ \citenamefont {Shen}}]{shen}%
  \BibitemOpen
  \bibfield  {author} {\bibinfo {author} {\bibfnamefont {W.~S.}\ \bibnamefont
  {Lee}}, \bibinfo {author} {\bibfnamefont {I.~M.}\ \bibnamefont {Vishik}},
  \bibinfo {author} {\bibfnamefont {K.}~\bibnamefont {Tanaka}}, \bibinfo
  {author} {\bibfnamefont {D.~H.}\ \bibnamefont {Lu}}, \bibinfo {author}
  {\bibfnamefont {T.}~\bibnamefont {Sasagawa}}, \bibinfo {author}
  {\bibfnamefont {N.}~\bibnamefont {Nagaosa}}, \bibinfo {author} {\bibfnamefont
  {T.~P.}\ \bibnamefont {Devereaux}}, \bibinfo {author} {\bibfnamefont
  {Z.}~\bibnamefont {Hussain}}, \ and\ \bibinfo {author} {\bibfnamefont
  {Z.-X.}\ \bibnamefont {Shen}},\ }\href@noop {} {\bibfield  {journal}
  {\bibinfo  {journal} {Nature}\ }\textbf {\bibinfo {volume} {450}},\ \bibinfo
  {pages} {81} (\bibinfo {year} {2007})}\BibitemShut {NoStop}%
\bibitem [{\citenamefont {Kondo}\ \emph {et~al.}(2009)\citenamefont {Kondo},
  \citenamefont {Takeuchi}, \citenamefont {Schmalian},\ and\ \citenamefont
  {Kaminski}}]{ref5}%
  \BibitemOpen
  \bibfield  {author} {\bibinfo {author} {\bibfnamefont {T.}~\bibnamefont
  {Kondo}}, \bibinfo {author} {\bibfnamefont {R.~K.~T.}\ \bibnamefont
  {Takeuchi}}, \bibinfo {author} {\bibfnamefont {J.}~\bibnamefont {Schmalian}},
  \ and\ \bibinfo {author} {\bibfnamefont {A.}~\bibnamefont {Kaminski}},\
  }\href@noop {} {\bibfield  {journal} {\bibinfo  {journal} {Nature (London)}\
  }\textbf {\bibinfo {volume} {457}},\ \bibinfo {pages} {296} (\bibinfo {year}
  {2009})}\BibitemShut {NoStop}%
\bibitem [{\citenamefont {DeGennes}(1999)}]{degennes}%
  \BibitemOpen
  \bibfield  {author} {\bibinfo {author} {\bibfnamefont {P.~G.}\ \bibnamefont
  {DeGennes}},\ }\href@noop {} {\emph {\bibinfo {title} {Superconductivity of
  Metals and Alloys}}},\ \bibinfo {edition} {3rd}\ ed.\ (\bibinfo  {publisher}
  {Westview Press},\ \bibinfo {year} {1999})\BibitemShut {NoStop}%
\bibitem [{\citenamefont {Kurosawa}\ \emph {et~al.}(2010)\citenamefont
  {Kurosawa}, \citenamefont {Yoneyama}, \citenamefont {Takano}, \citenamefont
  {Hagiwara}, \citenamefont {Inoue}, \citenamefont {Hagiwara}, \citenamefont
  {Kurusu}, \citenamefont {Takeyama}, \citenamefont {Momono}, \citenamefont
  {Oda},\ and\ \citenamefont {Ido}}]{kurosawa}%
  \BibitemOpen
  \bibfield  {author} {\bibinfo {author} {\bibfnamefont {T.}~\bibnamefont
  {Kurosawa}}, \bibinfo {author} {\bibfnamefont {T.}~\bibnamefont {Yoneyama}},
  \bibinfo {author} {\bibfnamefont {Y.}~\bibnamefont {Takano}}, \bibinfo
  {author} {\bibfnamefont {M.}~\bibnamefont {Hagiwara}}, \bibinfo {author}
  {\bibfnamefont {R.}~\bibnamefont {Inoue}}, \bibinfo {author} {\bibfnamefont
  {N.}~\bibnamefont {Hagiwara}}, \bibinfo {author} {\bibfnamefont
  {K.}~\bibnamefont {Kurusu}}, \bibinfo {author} {\bibfnamefont
  {K.}~\bibnamefont {Takeyama}}, \bibinfo {author} {\bibfnamefont
  {N.}~\bibnamefont {Momono}}, \bibinfo {author} {\bibfnamefont
  {M.}~\bibnamefont {Oda}}, \ and\ \bibinfo {author} {\bibfnamefont
  {M.}~\bibnamefont {Ido}},\ }\href@noop {} {\bibfield  {journal} {\bibinfo
  {journal} {Phys. Rev. B}\ }\textbf {\bibinfo {volume} {81}},\ \bibinfo
  {pages} {094519} (\bibinfo {year} {2010})}\BibitemShut {NoStop}%
\bibitem [{\citenamefont {Valla}\ \emph {et~al.}(2000)\citenamefont {Valla},
  \citenamefont {Fedorov}, \citenamefont {Johnson}, \citenamefont {Li},
  \citenamefont {Gu},\ and\ \citenamefont {Koshizuka}}]{arpesmfl2}%
  \BibitemOpen
  \bibfield  {author} {\bibinfo {author} {\bibfnamefont {T.}~\bibnamefont
  {Valla}}, \bibinfo {author} {\bibfnamefont {A.~V.}\ \bibnamefont {Fedorov}},
  \bibinfo {author} {\bibfnamefont {P.~D.}\ \bibnamefont {Johnson}}, \bibinfo
  {author} {\bibfnamefont {Q.}~\bibnamefont {Li}}, \bibinfo {author}
  {\bibfnamefont {G.~D.}\ \bibnamefont {Gu}}, \ and\ \bibinfo {author}
  {\bibfnamefont {N.}~\bibnamefont {Koshizuka}},\ }\href@noop {} {\bibfield
  {journal} {\bibinfo  {journal} {Phys. Rev. Lett.}\ }\textbf {\bibinfo
  {volume} {85(4)}} (\bibinfo {year} {2000})}\BibitemShut {NoStop}%
\bibitem [{\citenamefont {Jin}\ \emph {et~al.}(2011)\citenamefont {Jin},
  \citenamefont {Butch}, \citenamefont {Kirshenbaum}, \citenamefont
  {Paglione},\ and\ \citenamefont {Greene}}]{spinfluc}%
  \BibitemOpen
  \bibfield  {author} {\bibinfo {author} {\bibfnamefont {K.}~\bibnamefont
  {Jin}}, \bibinfo {author} {\bibfnamefont {N.~P.}\ \bibnamefont {Butch}},
  \bibinfo {author} {\bibfnamefont {K.}~\bibnamefont {Kirshenbaum}}, \bibinfo
  {author} {\bibfnamefont {J.}~\bibnamefont {Paglione}}, \ and\ \bibinfo
  {author} {\bibfnamefont {R.~L.}\ \bibnamefont {Greene}},\ }\href@noop {}
  {\bibfield  {journal} {\bibinfo  {journal} {Nature (London)}\ }\textbf
  {\bibinfo {volume} {476}},\ \bibinfo {pages} {73} (\bibinfo {year}
  {2011})}\BibitemShut {NoStop}%
\bibitem [{\citenamefont {Basov}\ and\ \citenamefont
  {Chubukov}(2011)}]{manifesto}%
  \BibitemOpen
  \bibfield  {author} {\bibinfo {author} {\bibfnamefont {D.~N.}\ \bibnamefont
  {Basov}}\ and\ \bibinfo {author} {\bibfnamefont {A.~V.}\ \bibnamefont
  {Chubukov}},\ }\href@noop {} {\bibfield  {journal} {\bibinfo  {journal}
  {Nature Physics}\ }\textbf {\bibinfo {volume} {7}},\ \bibinfo {pages} {272}
  (\bibinfo {year} {2011})}\BibitemShut {NoStop}%
\bibitem [{\citenamefont {Abrahams}\ and\ \citenamefont
  {Varma}(2000)}]{varma2}%
  \BibitemOpen
  \bibfield  {author} {\bibinfo {author} {\bibfnamefont {E.}~\bibnamefont
  {Abrahams}}\ and\ \bibinfo {author} {\bibfnamefont {C.~M.}\ \bibnamefont
  {Varma}},\ }\href@noop {} {\bibfield  {journal} {\bibinfo  {journal} {Proc.
  Natl. Acad. Sci. USA}\ }\textbf {\bibinfo {volume} {97}},\ \bibinfo {pages}
  {5714} (\bibinfo {year} {2000})}\BibitemShut {NoStop}%
\bibitem [{\citenamefont {Sachdev}(1999)}]{sachdev}%
  \BibitemOpen
  \bibfield  {author} {\bibinfo {author} {\bibfnamefont {S.}~\bibnamefont
  {Sachdev}},\ }\href@noop {} {\emph {\bibinfo {title} {Quantum Phase
  Transitions}}}\ (\bibinfo  {publisher} {Cambridge University Press},\
  \bibinfo {year} {1999})\BibitemShut {NoStop}%
\end{thebibliography}

%

\end{document}